\documentclass{article}
\usepackage[utf8]{inputenc}
\usepackage{natbib}
\usepackage{svg}

\usepackage[english]{babel}
\usepackage{multirow}
\usepackage{graphicx}

\usepackage[letterpaper,top=2cm,bottom=2cm,left=3cm,right=3cm,marginparwidth=1.75cm]{geometry}

\usepackage{amsmath}
\usepackage{graphicx}
\usepackage{colortbl}
\usepackage{xcolor}
\usepackage{pifont}
\usepackage{authblk}
\usepackage{comment}
\usepackage{amssymb}
\makeatletter
\newcommand{\printfnsymbol}[1]{%
  \textsuperscript{\@fnsymbol{#1}}%
}
\makeatother
\usepackage[colorlinks=true, allcolors=blue]{hyperref}
\newlength\savewidth\newcommand\shline{\noalign{\global\savewidth\arrayrulewidth\global\arrayrulewidth1pt}\hline\noalign{\global\arrayrulewidth\savewidth}}
\definecolor{baselinecolor}{gray}{.9}
\definecolor{darkgray}{gray}{.7}
\DeclareUnicodeCharacter{03B1}{\ensuremath{\alpha}}

\title{{\bf Prot42}\footnote{Prot42 is part of the Omics42 platform which includes also a family of genomic LMs and a family of chemical LMs named Gene42 and Chem42, respectively. Refer to \href{https://huggingface.co/spaces/inceptionai/Omics42}{\textbf{Omics42}} blog at \href{https://huggingface.co/inceptionai}{huggingface.co/inceptionai} for further details.} : a Novel Family of Protein Language Models for Target-aware Protein Binder Generation} 

\author[1]{Mohammad Amaan Sayeed}
\author[2]{Engin Tekin}
\author[1]{Maryam Nadeem}
\author[1]{Nancy A. ElNaker}
\author[1]{Aahan Singh}
\author[2]{Natalia Vassilieva}
\author[1]{Boulbaba {Ben Amor}\thanks{Corresponding author: Boulbaba Ben Amor \url{boulbaba.amor@inceptionai.ai}}}

\affil[1]{Inception Institute of Artificial Intelligence, Abu Dhabi, UAE.}
\affil[2]{Cerebras Systems, Sunnyvale, CA, USA.}
\date{}

\begin{document}
\maketitle

\begin{abstract}

Unlocking the next generation of biotechnology and therapeutic innovation demands overcoming the inherent complexity and resource-intensity of conventional protein engineering methods. Recent GenAI-powered computational techniques often rely on the availability of the target protein’s 3D structures and specific binding sites to generate high-affinity binders, constraints exhibited by models such as AlphaProteo and RFdiffusion. In this work, we explore the use of Protein Language Models (pLMs) for high-affinity binder generation. We introduce Prot42, a novel family of Protein Language Models (pLMs) pretrained on vast amounts of unlabeled protein sequences. By capturing deep evolutionary, structural, and functional insights through an advanced auto-regressive, decoder-only architecture inspired by breakthroughs in natural language processing, Prot42 dramatically expands the capabilities of computational protein design based on language only. Remarkably, our models handle sequences up to 8,192 amino acids, significantly surpassing standard limitations and enabling precise modeling of large proteins and complex multi-domain sequences. Demonstrating powerful practical applications, Prot42 excels in generating high-affinity protein binders and sequence-specific DNA-binding proteins. Our innovative models are publicly available, offering the scientific community an efficient and precise computational toolkit for rapid protein engineering. Explore our Foundation models at \href{https://huggingface.co/inceptionai}{huggingface.co/inceptionai}.

\begin{figure}[ht!]
    \centering
    \includegraphics[width=\linewidth]{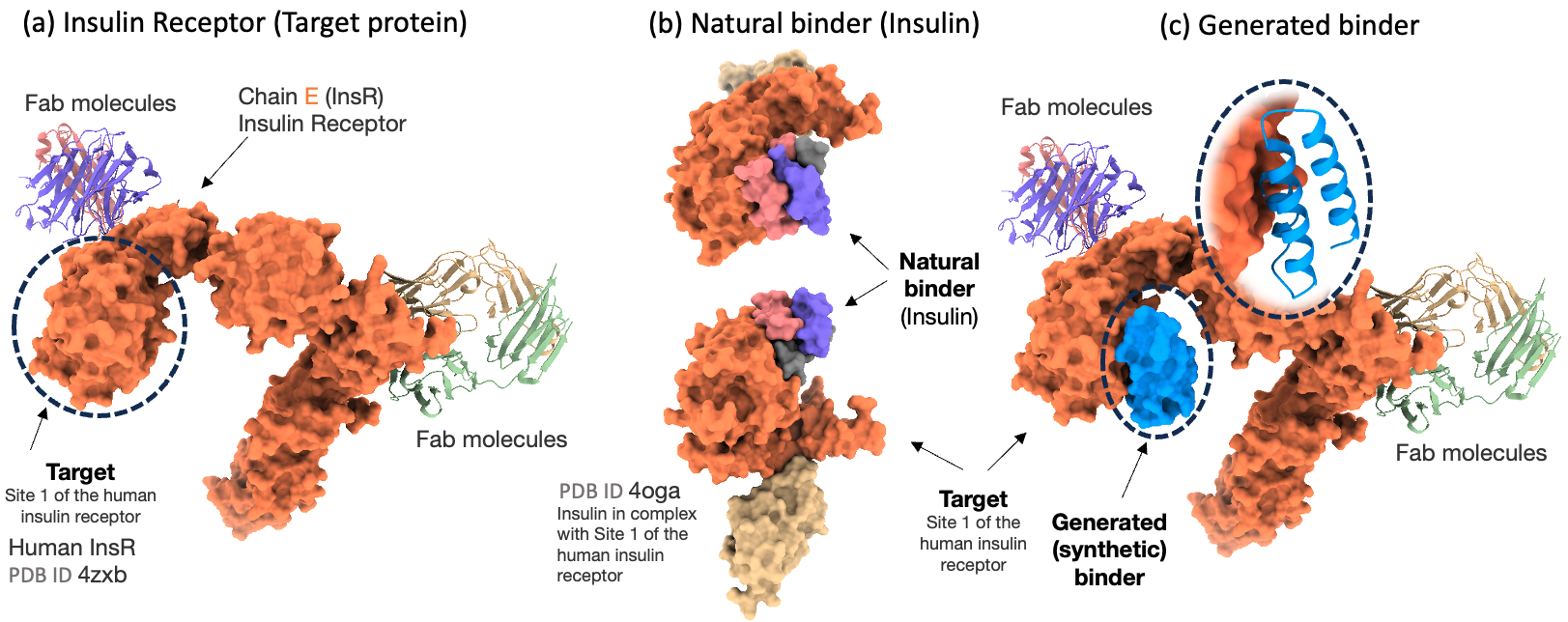}
    \caption{An example of generated protein binder to bind the Human InsR - Insulin Receptor protein (PDB ID 4zxb) in particular Site 1 (PDB ID 4oga) in panel (c) showing high binding affinity; (b) the natural binder (Ins - Insulin); and (a) the Insulin Receptor (Target protein). } 
    \label{fig:synthetic-insulin}
\end{figure}

\end{abstract}

%\tableofcontents

\section{Introduction}
\label{sec:Introduction}

Protein binders, including antibodies and engineered proteins, play pivotal roles in biotechnology and therapeutic applications ranging from diagnostics to targeted drug delivery~\cite{bradbury2011beyond,leader2008protein}. Traditional experimental methods for generating specific protein binders are resource-intensive and limited by combinatorial complexity. Structure-based computational approaches have emerged as promising alternatives, with notable advances including AlphaProteo~\cite{zambaldi2024novo} and RFdiffusion~\cite{watson2023novo}, which achieve high experimental success rates while requiring significantly fewer candidates to screen compared to previous methods.
Despite these advances, current approaches present key limitations: they fundamentally depend on target protein 3D structures and explicit binding site specifications, creating bottlenecks for targets with limited structural data. Protein language models (pLMs) offer promising alternatives by operating primarily at the sequence level~\cite{yang2022evaluating,elnaggar2021prottrans,rives2021biological}, leveraging large-scale unlabeled protein sequence data to learn representations that capture evolutionary relationships and structural properties~\cite{rao2021transformer}.
However, current pLMs face critical limitations in maximum sequence length and generative capabilities. Models like ESM-1b~\cite{rives2021biological} and ProtBert~\cite{elnaggar2021prottrans} demonstrate strong encoding abilities but lack native generative functionalities and typically constrain input to around 1,024 amino acids. This limits their effectiveness in modeling complex proteins and binding interfaces~\cite{xu2022peer}. Recent advances like Evo-2, a biological foundation model trained on 9.3 trillion DNA base pairs, have expanded capabilities with context windows up to one million tokens~\cite{brixi2024genome}, but primarily model DNA sequences rather than directly addressing protein-specific functionalities, highlighting the need for specialized protein-level generative models.

In this work, we introduce Prot42—a family of protein language models (pLMs) that harnesses the generative power of auto-regressive, decoder-only architectures, inspired by cutting-edge advancements in natural language processing, such as the LLaMA model \cite{touvron2023llama}. We pre-train two Prot42 variants, with 500 million and 1.1 billion parameters, initially supporting sequences up to 1,024 amino acids. Through continuous pretraining, we extend their context length to 8,192 residues, unlocking the ability to capture complex long-range dependencies essential for modeling large proteins, multidomain assemblies, and intricate molecular interactions. This enhanced representation power is pivotal for generating high-affinity protein binders, accelerating the discovery of novel biomolecular interactions. We detail our model architecture, pretraining methodologies, and our context-length scaling strategy, highlighting its impact on sequence modeling accuracy. Furthermore, we evaluate Prot42 using perplexity-based assessments, demonstrating its improved predictive performance across extended sequence contexts. To showcase its real-world applications, we present 1) Protein Binder Generation, with a particular focus on 2) Sequence-specific DNA-binding proteins. Our findings illustrate how Prot42’s advanced generative capabilities redefine the frontiers of computational protein design, enabling rapid, precise, and scalable protein engineering. As an initial demonstration of Prot42’s design power, Figure \ref{fig:synthetic-insulin} presents a high-affinity binder, computationally generated to target the α -subunit of the Insulin Receptor (InsR), compared to the natural insulin binder (PDB ID 4oga).

\section{Related Works}
\label{sec:SoTA}

Protein language models (pLMs) have emerged as powerful tools in computational biology by learning rich representations directly from extensive unlabeled protein sequence databases. This approach effectively bridges the gap between the large number of known protein sequences and the relatively small subset ($<$0.3\%) with experimentally verified functions~\cite{yang2022evaluating}. pLMs capture intricate evolutionary and biochemical patterns, surpassing traditional features based on physicochemical or statistical analyzes~\cite{Chou2001,Dubchak1995,Shen2007,Altschul1997,Zou2011}. Consequently, they have significantly improved various tasks, including annotation of protein functions, structural prediction, and novel sequence generation~\cite{yang2022evaluating}. Early deep learning models adapted natural language processing (NLP) frameworks such as word2vec and doc2vec to protein sequences, effectively capturing evolutionary and functional motifs~\cite{Mikolov2013,Le2014,Yang2018,Asgari2015,Bepler2019,Rao2019}. These initial models laid the foundation for deeper and more sophisticated architectures pre-trained on larger datasets, leading to substantial improvements in protein representation quality. Notable early examples include UniRep~\cite{Alley2019} and ProtXLNet~\cite{Elnaggar2021}, which leveraged autoregressive next amino acid prediction strategies.

\subsection*{Protein Language Models}
Transformer-based architectures marked a significant leap forward. Models such as TAPE Transformer~\cite{Rao2019}, ProtBert, ProtAlbert, ProtElectra, and ProtT5~\cite{elnaggar2021prottrans} have adopted masked language modeling (MLM) approaches, significantly advancing protein representation learning. ProtBert, a key model from the ProtTrans family, utilizes a BERT-adapted bidirectional transformer encoder architecture, consisting of 30 layers and 16 attention heads per layer with approximately 420 million parameters. ProtBert was pre-trained on the massive BFD dataset, containing around 2.1 billion protein sequences, significantly enriching its capability to represent evolutionary variations. ProtBert has shown robust performance in various protein prediction tasks, although slightly behind ESM-1b, with a mean reciprocal rank (MRR) around 0.23~\cite{elnaggar2021prottrans,edera2022enhanced,xu2022peer}. ESM-1 (Evolutionary Scale Modeling) introduced transformer-based pLMs trained using masked language modeling on UniRef50 database, a dataset comprising clusters of UniProt sequences at 50\% sequence identity~\cite{rives2021biological}. ESM-1b, a prominent variant, utilizes a deep transformer architecture consisting of 34 self-attention layers and approximately 650 million parameters, achieving state-of-the-art performance in protein function and structure prediction tasks, including remote homology detection and mutational effect predictions. When augmented with auxiliary contact prediction tasks, ESM-1b notably achieves MRR of approximately 0.517~\cite{xu2022peer,rives2021biological,rao2021transformer}. ESM-2 expanded upon ESM-1 by increasing model complexity and refining training protocols, enhancing its representational capabilities. It utilized larger datasets and more sophisticated training regimes, achieving superior results on protein functional annotation and structure prediction tasks compared to previous versions~\cite{lin2023esm2}. The recently introduced ESM-3 \cite{esm3} significantly scales the model parameters and introduces advanced architectural innovations designed to capture intricate structural motifs and long-range dependencies more effectively. These improvements solidify its position as a state-of-the-art transformer-based PLM, offering enhanced capabilities in structural and functional protein modeling tasks~\cite{lin2023esm2}. Structural information, including multiple sequence alignments (MSA), three-dimensional (3D) structures, and surface features, has also been integrated into pLMs, enhancing their representational learning power ~\cite{Rao2021,Biswas2021,Sturmfels2020,Senior2020,Jumper2021,Gainza2020,Sverrisson2021,Kipf2020}. However, sequence-based methods remain dominant because of the significantly larger availability of sequence data compared to structural data. Benchmarks such as CASP~\cite{Moult2005}, CAFA~\cite{Radivojac2013}, TAPE~\cite{Rao2019}, FLIP~\cite{Dallago2021}, TDA~\cite{Chen2022}, ATOM3D~\cite{Townshend2021}, and PEER~\cite{xu2022peer} systematically evaluate these models, consistently demonstrating the superiority of transformer-based pLMs.

\subsection*{Computational Design of Protein Binders}

Protein binders, including antibodies and engineered proteins, play pivotal roles in biotechnology and therapeutic applications ranging from diagnostics and imaging to targeted drug delivery~\cite{bradbury2011beyond,leader2008protein}. Traditionally, generating highly specific protein binders relies extensively on experimental techniques such as phage display and directed evolution~\cite{smith1985filamentous,packer2015directed}. Despite their efficacy, these methods are resource-intensive, time-consuming, and limited by the combinatorial complexity inherent in protein sequences. 

While recent advancements in protein binder design have leveraged both structure-based and sequence-based approaches, each method remains constrained in several significant ways. Structure-based methods such as AlphaProteo \cite{zambaldi2024novo}, RFDiffusion \cite{watson2023rf}, and MASIF-Seed \cite{gainza2020deciphering} rely heavily on extensive high-resolution structural data, restricting their applicability to protein targets with well-defined three-dimensional conformations. For instance, AlphaProteo demonstrates experimental success rates of 9-88\% across diverse targets, significantly outperforming previous methods. However, AlphaProteo requires screening between 54-172 designs per target, with previous similar approaches needing thousands to hundreds of thousands of designs to achieve comparable results. Similarly, RFdiffusion~\cite{watson2023novo} employs diffusion models operating in 3D structural space to create novel protein binders, achieving success rates of 0-33\% across various targets while typically screening 95-15,000 candidates per target. Both approaches have yielded impressive results for therapeutic candidates, but they fundamentally rely on the availability of target protein 3D structures and explicit binding site specifications. This structural dependency and computational intensity create bottlenecks in rapidly designing binders for targets with limited or no structural data. 

Conversely, recent sequence-based models, including PepMLM \cite{ferruz2022protGPT2}, ProGen2 \cite{nijkamp2023progen}, and moPPIt \cite{chen2024moppit}, have demonstrated promising generative capabilities but remain primarily effective for shorter peptides and struggle to generalize to larger, therapeutically relevant protein binders \cite{chen2024moppit}. Additionally, these methods typically require auxiliary task-specific training objectives or manual curation of functional motifs, further limiting their applicability. While moPPIt advances epitope-specific binding through its multi-objective optimization approach, it remains computationally unvalidated for disordered targets and would benefit from experimental confirmation of its predicted binding interactions.

Despite these advances, there remains a significant gap in the field for protein language models that combine extended context length capabilities with true generative power, especially for designing full-length protein binders with complex binding interfaces and long-range dependencies. This gap motivates our development of Prot42, which specifically addresses these limitations through its architecture and training methodology.

\subsection*{Sequence-specific DNA-binding Proteins Design}

 Sequence-specific DNA-binding proteins are crucial regulators of gene expression in all organisms. Transcription factors (TF) represent a primary example of this class, binding to specific DNA sequences - typically within promoter or enhancer regions - to regulate the transcription of target genes \citep{Spa24}. By recruiting or blocking RNA polymerase and interacting with various cofactors, TFs precisely coordinate when and where genes are activated or silenced, thus directing essential cellular functions and establishing cell identity. Numerous vital biological processes, such as development, cell cycle control, and responses to environmental stress, rely on sophisticated networks of DNA-binding proteins that often act collaboratively. Disruptions or mutations in these proteins can lead to abnormal gene expression, contributing to disease development, including cancers driven by misregulated TFs. In addition, other sequence-specific DNA-binding proteins, such as bacterial restriction enzymes (which recognize and cleave foreign DNA sequences) and DNA repair or recombination proteins (targeting particular DNA motifs), play crucial roles in maintaining genomic integrity. Collectively, sequence-specific DNA binding proteins interpret genomic regulatory information, ensuring precise gene activation and maintaining normal cellular function \citep{Spa24}. Given their fundamental biological significance, computational methods for identifying and characterizing DNA-binding proteins have attracted extensive research interest. Historically, traditional machine learning approaches employed features such as amino acid composition, conserved motifs, and evolutionary profiles, establishing foundational insights. However, deep learning techniques, particularly convolutional neural networks (CNNs) and recurrent neural networks (RNNs), which take advantage of extensive and comprehensive datasets, have significantly improved the accuracy of DNA-binding residue and motif predictions \citep{IEE24,KMLX24}. Integrating protein language models (pLMs), such as ProtTrans, with multi-window CNN architectures has further improved performance, capturing intrinsic biochemical properties and sequence motifs essential for DNA-binding recognition, as demonstrated by a remarkable area under the ROC curve (AUC) of 0.89 \citep{LLC+24}. Furthermore, graph-based neural networks that integrate three-dimensional structural contexts via residue contact maps and spatial graphs have further elevated predictive capabilities. These methods, supported by breakthroughs in protein structure prediction such as AlphaFold2, now offer highly accurate computational tools for genome annotation, elucidation of gene regulatory networks, and the development of targeted gene editing technologies. Advances in protein foundational models not only bolster predictive accuracy, but also open transformative possibilities for the generation of novel DNA-binding proteins. Recent computational design methodologies, exemplified by the work of Glasscock et al., \citep{GPM+24} utilize powerful generative capabilities inherent in foundational protein models to engineer novel proteins that recognize specific DNA sequences through major groove interactions. These designed proteins exhibited precise target sequence specificity and affinities in the nanomolar range (30–100 nM). Structural validation through crystal structures underscored the high accuracy and reliability of these computational models. Importantly, these newly generated proteins demonstrated effective modulation of transcriptional activity in both \textit{Escherichia coli} and mammalian cells, underscoring their practical potential. This integration of protein foundational models with computational design significantly advances our ability to create sequence-specific DNA-binding proteins that are easily deployable, highlighting their vast applicability in gene regulation and genome-editing applications \citep{GPM+24}. 

\section{Methodology}
\label{sec:Methodology}
Leveraging Prot42, we demonstrate for the first time an instruction-tuning approach to generate high-affinity, long length protein binders directly from target sequences alone, without incorporating structural information or additional auxiliary training objectives. By utilizing a novel multimodal strategy that integrates Gene42's genomic embeddings with Prot42's protein embeddings, we further extend our capabilities to generate sequence-specific DNA-binding proteins. This approach eliminates dependence on explicit structural constraints or predefined binding motifs, highlighting the intrinsic capacity of computational models to design functional protein binders across diverse molecular contexts.
\subsection{Data Preparation}
UniRef50 comprises 63.2 million amino acid sequences, which are tokenized using a vocabulary of 20 standard amino acids. To account for any uncommon or ambiguous residues, we use X token representing any amino acid. Each sequence is processed with a maximum context length of 1,024 tokens, and sequences exceeding this limit are excluded, resulting in a filtered dataset of 57.1 million sequences with an initial packing density of 27\%. To optimize data utilization and improve computational efficiency, we employ variable sequence length (VSL) packing, which maximizes token occupancy within the fixed context length. This approach significantly enhances packing density, reducing the dataset to 16.2 million packed sequences while achieving a packing efficiency of 96\%. This refined dataset ensures a more efficient use of computational resources while preserving sequence diversity and integrity.

\begin{table}[ht!]
\centering
\scriptsize
\addtolength{\tabcolsep}{0pt}
\def\arraystretch{1.2}
    \begin{tabular}{l| c  c |  c c c}

            \textbf{Model} &  \textbf{Prot42-B} &  \textbf{Prot42-L} & \textbf{Prot42-L 2K} &  \textbf{Prot42-L 4K} &  \textbf{Prot42-L 8K} \\
            \shline
            \# of parameters & 500M & 1.1B & 1.1B & 1.1B & 1.1B\\
            Sequence Length & 1024 & 1024 & 2048 & 4096 & 8192 \\
            Effective Length & 983 & 983 & 1331 & 2662 & 5324 \\
            Tokens per Batch & 1M & 1M & 1M & 1M & 1M \\
            Batch size & 1024 & 1024 & 756 & 378 & 189\\
            Base Frequency & 10k & 10k & 10k & 10k & 10k \\
            Hidden size & 1,408 & 2,048 & 2,048 & 2,048 & 2,048 \\
            \# of hidden layers & 16 & 24 & 24 & 24 & 24 \\
            \# of attention heads & 16 & 32 & 32 & 32 & 32 \\
            Transformer FFN Dim. & 5632 & 5440 & 5440 & 5440 & 5440 \\
        \hline
            Optimizer & AdamW & AdamW & AdamW & AdamW & AdamW \\
            Betas & 0.9, 0.95 & 0.9, 0.95 & 0.9, 0.95 & 0.9, 0.95 & 0.9, 0.95\\
            Eps & 1e-8 & 1e-8 & 1e-8 & 1e-8 & 1e-8\\
            Weight Decay & 0.1 & 0.1 & 0.1 & 0.1 & 0.1\\
            Max grad norm & 1 & 1 & 1 & 1 & 1\\
            Learning rate (Linear) & 0 to 4.8e-4 & 0 to 4.8e-4 & - & - & -\\
            Iterations (Linear) & 0 to 2000 & 0 to 3950 & - & - & - \\
            Learning rate (Cosine) & 4.8e-4 to 4.8e-5 & 4e-4 to 1e-5 & 4e-4 to 1e-5 & 4e-4 to 1e-5 & 4e-4 to 1e-5 \\
            Iterations (Cosine) & 2000 to 126000 & 3950 to 237000 & 1860 to 3726 & 3726 to 4305 & 4305 to 4472 \\
        
    \end{tabular}
    \caption{\textbf{Hyperparameters used for pretraining (left) and continuous pretraining (right)} of the Prot42 models.}
    \label{table:Prot42-p-models-training-hyperparams}
\end{table}

\subsection{Model Architecture and Pretraining}
\label{sec:Architecture}
Prot42 is an autoregressive transformer decoder model following the LLaMA architecture \cite{touvron2023llama}. We pre-train two model variants with parameter counts of 500 million (500M) and 1.1 billion (1.1B), each utilizing a maximum sequence length (MSL) of 1024 tokens. We denote the 500M-parameter model as \textbf{Prot42-B} (base) and the 1.1B-parameter model as \textbf{Prot42-L} (large).

To optimize hyperparameter selection for training, we adopt maximal update parametrization ($\mu$P) \cite{mup-mutransfer}. Our hyperparameter tuning process is carried out using a smaller 81M parameter proxy model, from which optimal parameters are identified and subsequently transferred to the 500M and 1.1B target models by $\mu$ transfer \cite{mup-mutransfer}. During pre-training we keep number of tokens per batch constant at 1 million, apply peak learning rate of 4.8e4 with 10x cosine decay and initial linear warm-up schedule. We utilize Cerebras CS-2 for our training runs. The Cerebras CS-2 system is an AI accelerator that features 850,000 AI optimized compute cores, 40GB of on-chip SRAM, 20 PB/s memory bandwidth, and 220 PB/s interconnect \cite{cs2specs}. Table \ref{table:Prot42-p-models-training-hyperparams} shows our model architecture and pre-training configuration in detail.

\subsubsection*{Continuous Pretraining for a Larger Context Length}
The Prot42-L model was further fine-tuned to accommodate extended context lengths, achieved through a gradually increased MSL. Initially, training started with an MSL of 1,024 tokens, which constitute 10\% of the total training steps. This phase was followed by a 10x cosine-decay schedule, during which the context length was progressively increased to 2,048, 4,096, and 8,192 tokens. Throughout these stages, the number of non-padding tokens per batch was maintained at a constant 1 million tokens. Table \ref{table:context_extension_dataset} presents the dataset splits used for context extension, while the hyper-parameter details are provided in Table \ref{table:Prot42-p-models-training-hyperparams}.

\begin{table}[ht!]
\centering
\footnotesize
\begin{tabular}{l|l|l|l|l|l|l|l}
%\hline
\textbf{MSL} & \textbf{\# Samples} & \textbf{Val Size} & \textbf{Train Size} & \textbf{BS} & \textbf{Steps} & \textbf{Effective \# Tokens}  & \textbf{Effective MSL} \\
\hline
1024 & 8,257,804 & 639,244 & 7,618,560 & 4,096 & 1,860 & 1,872,337,306 & 245\\
%\hline
2048 & 1,485,108 & 74,255 & 1,410,853 & 756 & 1,866 & 1,878,127,514 & 1331\\
%\hline
4096 & 230,559 & 11,527 & 219,032 & 378 & 579 & 583,150,797 & 2662\\
%\hline
8192 & 35,161 & 3,516 & 31,645 & 189 & 167 & 168,503,296 & 5324\\
%\hline

\end{tabular}
\caption{Context Length (MSL: Maximum Sequence Length) Extension Dataset of Prot42.}
\label{table:context_extension_dataset}
\end{table}

\subsubsection*{Model Evaluation using Validation Perplexity}
\label{sec:perplexity}
\begin{figure}[ht!]
    \centering
    \includegraphics[width=0.8\linewidth]{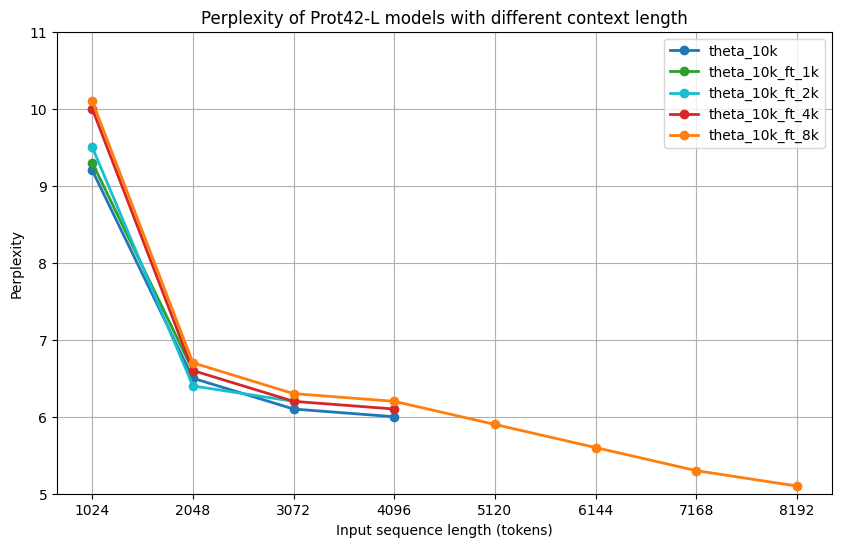}
   \caption{Validation Perplexity (PPL) of Prot42-L models with different context length on the validation dataset. The input sequence lengths are varied from 1k to 8k.}
    \label{fig:Figure 2}
\end{figure}

To evaluate the performance of our model before downstream task validation, we employed perplexity (PPL), a standard metric for evaluating autoregressive language models. For a tokenized amino acid sequence $X = {x_0, x_1, \ldots, x_n}$, PPL is defined as the exponentiated average negative log-likelihood: $\text{PPL}(X) = \exp{-\frac{1}{t} \sum_{i} \log p_\theta (x_i|x_{<i})}$, where $\log p_\theta$ is the log-likelihood of the $i$th token conditioned on preceding tokens $x_{<i}$, and $t$ is the context length.
Figure \ref{fig:Figure 2} shows the PPL of Prot42-L models with different context lengths on the validation dataset. We varied input sequence lengths from 1k to 8k to thoroughly test model capabilities. All models initially show relatively high perplexity (9-10) at 1024 tokens, with a substantial improvement at 2048 tokens, decreasing to approximately 6.5.

% \label{sec:perplexity}

Our base model and those fine-tuned for shorter contexts show comparable performance patterns up to their respective maximum context lengths. The 8k context model demonstrates particularly interesting behavior – while it shows slightly higher perplexity in mid-range sequences (2048-4096 tokens), it uniquely processes sequences up to 8192 tokens, reaching its lowest perplexity of 5.1 at maximum length.
This declining perplexity curve beyond 4096 tokens indicates that our 8k model effectively leverages the expanded context window to capture long-range dependencies in protein sequences. Such capability is crucial for accurately modeling multidomain proteins and protein complexes that frequently exceed standard 1k or 2k residue thresholds of typical protein language models.
Our extended context window represents a significant advancement in protein sequence modeling, enabling more accurate representation of complex proteins and protein-protein interactions essential for effective protein binder generation.

\subsubsection*{Embeddings Evaluation}

Proteins orchestrate a wide range of cellular processes, with subcellular localization serving as a key determinant of function, interaction networks, and regulatory mechanisms. Accurate prediction of a protein’s localization is critical for unraveling its biological role and guiding applications in drug discovery, synthetic biology, and functional annotation. This section evaluates the representational power of the embeddings generated by Prot42-L, assessing their effectiveness in capturing biologically meaningful protein localization patterns across cellular compartments. Traditional protein prediction methodologies have predominantly relied on Multiple Sequence Alignments (MSAs) to infer functional and structural information. However, recent advances have highlighted that embedding-based predictions using protein language models (pLMs), trained solely on amino acid sequences, often meet or exceed the performance of state-of-the-art MSA-based methods in various prediction tasks \cite{schmirler2024fine}.

\begin{figure}[ht!]
    \centering
    \includegraphics[width=1\textwidth]{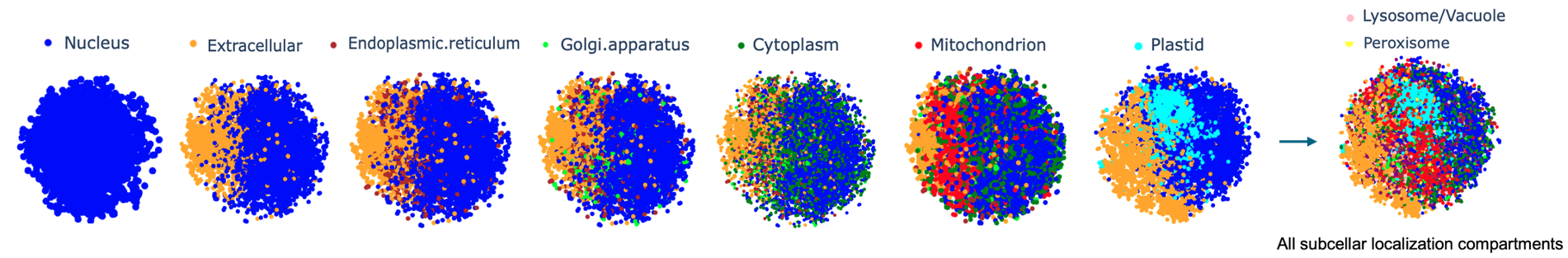}
    \caption{t-SNE visualization of \textbf{Prot42-L} protein embeddings across 10 subcellular localization compartments. Proteins cluster based on their localization as captured by \textbf{Prot42-L} embeddings.}
    \label{fig:figure1}
\end{figure}

The subcellular localization dataset comprises \textbf{PEER benchmark database} and \textbf{UniProt} annotated proteins, covering \textbf{10 subcellular compartments}: nucleus, cytoplasm, mitochondrion, endoplasmic reticulum, Golgi apparatus, lysosome/vacuole, peroxisome, extracellular space, peroxisome, and plasma membrane (more details are given in the appendix). Each protein sequence was represented as a high-dimensional vector of size \(32 \times 2048\). To ensure an effective global representation that captures comprehensive localization-specific contexts, we computed the mean token level of the embeddings of the \textbf{Prot42-L} model throughout the protein sequence. To visually evaluate the quality of the embeddings and compartmental differentiation, we applied stochastic neighbor embedding distributed by t \textbf{ (t-SNE)} to reduce the dimensionality, allowing a clear visualization of protein groups based on subcellular locations (Figure \ref{fig:figure1}). The t-SNE plots demonstrate distinct clusters corresponding to major cellular compartments such as the nucleus, mitochondrion, and extracellular regions, emphasizing the model's capability to discern proteins based on intrinsic localization-specific features. Meanwhile, proteins associated with the cytoplasm and the Golgi apparatus demonstrated partially overlapping distributions, reflecting the dynamic interactions and shared functional roles between these compartments. The observed clusters align closely with the established cellular topography as previously described by \cite{Chen2014}, confirming the effectiveness of the model in capturing the representations of subcellular localization proteins.

The embeddings learned by \textbf{Prot42-L} not only differentiate proteins according to their subcellular compartments, but also provide a versatile foundation for downstream predictive tasks. These high-dimensional representations can easily be integrated into various neural network architectures, including feedforward artificial neural networks (ANNs), convolutional neural networks (CNNs), and more sophisticated architectures that utilize structural and functional protein features. Further improvements could involve fine-tuning these broadly trained protein language models on task-specific datasets, or training specialized models tailored to protein families, such as antibodies, to enhance the accuracy and specificity of protein annotation. The ability of \textbf{Prot42-L} to encode subcellular localization through embeddings has significant implications for downstream applications in protein function prediction, drug target identification, and synthetic biology. Using these learned representations, our findings highlight the quality of \textbf{Prot42-L} embeddings, paving the way for advances in protein localization prediction, targeted drug development, synthetic biology, and large-scale functional annotation initiatives and downstream tasks. 

\subsection{Protein Binder Generation}
\label{sec:Protein-Binders}

Proteins do not function in isolation; they interact, bind, and form complex networks that drive cellular processes. Designing proteins that can specifically bind to a target protein is a fundamental problem in molecular biology, with applications ranging from therapeutics to synthetic biology. Let \( \mathbf{x} = (\mathbf{x}_1, \mathbf{x}_2, \dots, \mathbf{x}_n) \in \mathcal{X} \) represent the amino acid sequence of a \textbf{target protein}, where \( \mathbf{x}_i \) is the \( i \)-th amino acid in the sequence, and \( \mathbf{y} = (\mathbf{y}_1, \mathbf{y}_2, \dots, \mathbf{y}_m) \in \mathcal{Y} \) represent the amino acid sequence of a \textbf{binding protein} that binds to the target protein. The goal is to model the conditional probability distribution \( p(\mathbf{y} | \mathbf{x}) \), which represents the probability of the binding protein sequence \( \mathbf{y} \) given the target protein sequence \( \mathbf{x} \). To achieve this, we use a sequence-to-sequence model inspired by machine translation, trained on a dataset \( \mathcal{B} = \{(\mathbf{x}_i, \mathbf{y}_i)\} \) consisting of pairs of target and binder sequences sampled from known protein interactions. Each pair \( (\mathbf{x}, \mathbf{y}) \in \mathcal{B} \) is structured as \( \mathbf{s} = (\mathbf{x}_1, \mathbf{x}_2, \dots, \mathbf{x}_n, \text{SEP}, \mathbf{y}_1, \mathbf{y}_2, \dots, \mathbf{y}_m) \), where \( \text{SEP} \) is a special separator token to distinguish between the target and binding protein sequences. During training, we optimize the autoregressive loss function:

\[
\mathcal{L}(\boldsymbol{\theta}) = -\sum_{i=n+2}^{n+m+1} \log p_{\boldsymbol{\theta}}(\mathbf{s}_i | \mathbf{s}_{<i}),
\]

where \( \boldsymbol{\theta} \) represents the model parameters, and the summation runs over the entire sequence \( \mathbf{s} \) from both the target and binding protein sequences. After training, the model can generate binding protein sequences for a given target protein sequence. The process begins by conditioning the model on the target protein sequence \( \mathbf{x} \), followed by the separator token \( \text{SEP} \), and an initial methionine residue ('M') to prime the generation process, as protein sequences typically begin with methionine in the training distribution. The binding protein sequence is generated autoregressively as:

\[
p(\mathbf{y} | \mathbf{x}) = \prod_{i=1}^{m} p(\mathbf{y}_i | \mathbf{x}, \text{SEP}, \mathbf{y}_{<i}),
\]

where \( \mathbf{y} = (\mathbf{y}_1, \mathbf{y}_2, \dots, \mathbf{y}_m) \) is the binding protein sequence being generated, and \( \mathbf{y}_{<i} \) represents the sequence of previously generated binding proteins. To ensure the generation of diverse and high-quality binding proteins, we use a stochastic sampling approach incorporating temperature scaling, nucleus sampling (top-\( p \)), and top-\( k \) filtering. The probability of selecting the \( i \)-th token during sampling is given by:

\[
p_{\text{sample}}(\mathbf{y}_i | \mathbf{x}, \text{SEP}, \mathbf{y}_{<i}) \propto \begin{cases}
\frac{\exp(z_i / T)}{\sum_{j \in V'} \exp(z_j / T)} & \text{if } i \in V' \\
0 & \text{otherwise}
\end{cases}
\]

where \( T \) is the temperature parameter controlling randomness, \( z_i \) is the logit for token \( \mathbf{y}_i \), and \( V' \) is the subset of vocabulary tokens determined by the top-\( k \) and nucleus sampling thresholds. This sampling strategy helps balance exploration and quality, resulting in high-quality, diverse binding protein sequences.
Thus, the entire protein binding sequence generation process can be represented as follows:

\[
\hat{\mathbf{Y}} = M_p(\mathbf{X}, \text{SEP}, \mathbf{M}),
\]

where \( \hat{\mathbf{Y}} \) is the generated binding protein sequence, \( \mathbf{X} = (\mathbf{x}_1, \mathbf{x}_2, \dots, \mathbf{x}_n) \) is the input target protein sequence. The model ensures that the generated sequence \( \hat{\mathbf{Y}} = (\mathbf{y}_1, \mathbf{y}_2, \dots, \mathbf{y}_m) \) depends on both the target protein sequence and previously generated tokens. 

\begin{figure}[ht!]
    \centering
    \includegraphics[width=\linewidth]{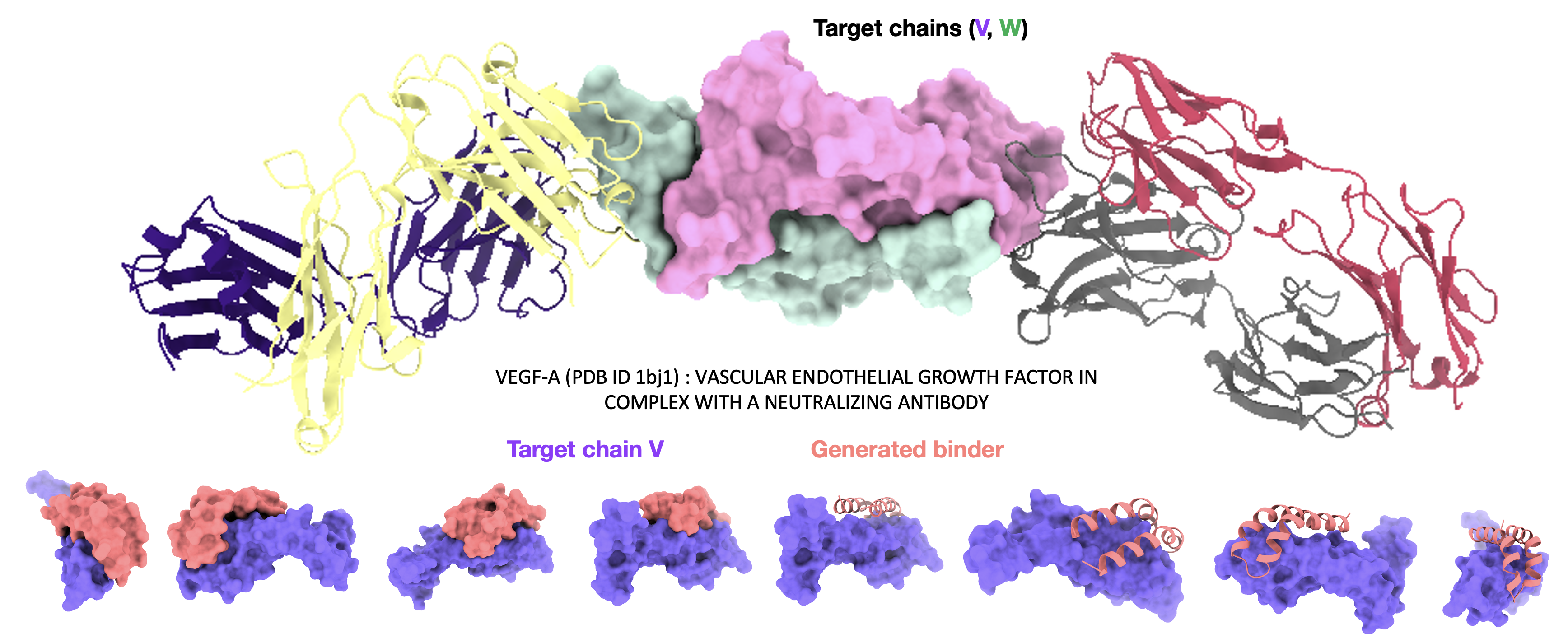}
    \caption{Illustration of the protein binding sequence generation process. Top: Target protein Vascular endothelial growth factor VEGF-A (PDB ID 1bj1) in complex with neutralizing antibody. Bottom: Multiple views and representations of a binding protein (in blue) generated by our Prot42 model for Chain V of VEGF-A. The model conditions on the target sequence $\mathbf{X}$ (VEGF-A), followed by $\text{SEP}$ token and initial methionine residue, generating binding sequence $\hat{\mathbf{Y}}$ autoregressively using $p(\mathbf{y} | \mathbf{x})$. The generated binder demonstrates a \textbf{$K_d$ of 4.2nM}, showcasing the model's capability to design proteins with specific binding properties. }
    \label{fig:1bj1}
\end{figure}

Figure \ref{fig:1bj1} provides an example of this process, illustrating the binding interaction between a generated binder and its target. In this case, the target protein \( \mathbf{X} \) corresponds to VEGF-A, and the generated binder \( \hat{\mathbf{Y}} \) is designed for Chain V. The model learns \( p(\mathbf{y} | \mathbf{x}) \) and generates \( \hat{\mathbf{Y}} \) autoregressively, following the structured input \( \mathbf{s} = (\mathbf{x}, \text{SEP}, \mathbf{y}) \). During inference, it conditions on \( \mathbf{X} \), initializes decoding with \( \text{SEP} \) and methionine (\(\mathbf{M}\)), and samples \( \mathbf{y}_i \) using:
$p_{\text{sample}}(\mathbf{y}_i | \mathbf{x}, \text{SEP}, \mathbf{y}_{<i})$.
The structural views in Figure~\ref{fig:1bj1} show 
\(\hat{\mathbf{Y}}\) binding to \(\mathbf{X}\), with dissociation constant 
\(K_d(\hat{\mathbf{Y}}, \mathbf{X})\) quantifying interaction strength. More results on the generation procedure will be discussed in Section~\ref{sec:protein_binders_generation}.

\subsection{DNA Sequence-Specific Binders Generation}

Beyond protein-protein interactions, designing proteins that bind to specific DNA sequences opens new frontiers in gene regulation and genome engineering. Given a dataset $\mathcal{D}$ of DNA-protein sequence pairs, our goal is to generate protein sequences $\hat{X}_p$ that interact effectively with a target DNA. Let $(X_d^{(1)}, X_d^{(2)}, X_p) \in \mathcal{D}$ represent a DNA-protein pair, where $X_d^{(1)} = (d_1^{(1)}, d_2^{(1)}, \dots, d_n^{(1)})$ and $X_d^{(2)} = (d_1^{(2)}, d_2^{(2)}, \dots, d_n^{(2)})$ are two DNA sequences, each of length $n$, belonging to a predefined vocabulary $\mathcal{V}_d$, and $X_p = (p_1, p_2, \dots, p_m)$ is a sequence of $m$ amino acid tokens representing the protein. The GFM model $M_d$ encodes both DNA sequences $X_d^{(1)}$ and $X_d^{(2)}$ into sequences of latent embeddings $\mathbf{H}_d^{(1)} = (h_1^{(1)}, h_2^{(1)}, \dots, h_n^{(1)})$ and $\mathbf{H}_d^{(2)} = (h_1^{(2)}, h_2^{(2)}, \dots, h_n^{(2)})$, where each $h_t^{(i)} \in \mathbb{R}^{1408}$ represents the hidden state of the genomic model for DNA sequence $X_d^{(i)}$.

The protein model $M_p$ encodes $X_p$ into a sequence of hidden representations $\mathbf{E}_p = (e_1, e_2, \dots, e_m)$, where each $e_i \in \mathbb{R}^{2048}$ captures structural and functional properties of the protein. 

To incorporate DNA context into protein sequence generation, DNA embeddings from both sequences are projected into the hidden dimension of the protein model using learnable transformations:
\[
\mathbf{H}_d^{(1)'} = \theta_{d_1} \mathbf{H}_d^{(1)}, \quad \mathbf{H}_d^{(2)'} = \theta_{d_2} \mathbf{H}_d^{(2)},
\]
where $\theta_{d_1} \in \mathbb{R}^{1408 \times 2048}$ and $\theta_{d_2} \in \mathbb{R}^{1408 \times 2048}$. These transformed DNA embeddings are integrated into the protein representations using a cross-attention mechanism. Attention scores are computed as $A = \text{Softmax} \left( \frac{QK^\top}{\sqrt{2048}} \right),$
where the query, key, and value projections are defined as;
\[
Q = \theta_q \left( \mathbf{H}_d^{(1)} \oplus \mathbf{H}_d^{(2)} \right)
, \quad K = \theta_k \mathbf{E}_p, \quad V = \theta_v \mathbf{E}_p,
\]
with learnable weight matrices $\theta_q, \theta_k, \theta_v \in \mathbb{R}^{2048 \times 2048}$. The resulting DNA-conditioned protein representations are given by $\mathbf{C}_p = A V$, which are combined with the original protein embeddings as $\mathbf{E}_p' = \mathbf{E}_p + \mathbf{C}_p$. These transformed representations $\mathbf{E}_p'$ are then passed through the protein decoder layers.

The generation of the protein sequence follows an autoregressive process in which each token $\hat{p}_t$ is predicted based on previously generated tokens and DNA-informed features. The probability distribution over the protein vocabulary $\mathcal{V}_p$ is given by
$p_t = \text{Softmax}(\theta_h e_t' + \theta_c C_t),$
where $\theta_h, \theta_c$ are learnable parameters, and the next protein token is selected as
$\hat{p}_t = \arg \max_j (p_{t,j}).$
This process continues until a termination token is reached. Thus, the entire DNA-to-protein sequence generation process can be represented as;

\[
\hat{X}_p = M_p(\mathbf{H}_d^{(1)'}, \mathbf{H}_d^{(2)'})
\]

\begin{figure}[ht!]
    \centering
    \includegraphics[width=\linewidth]{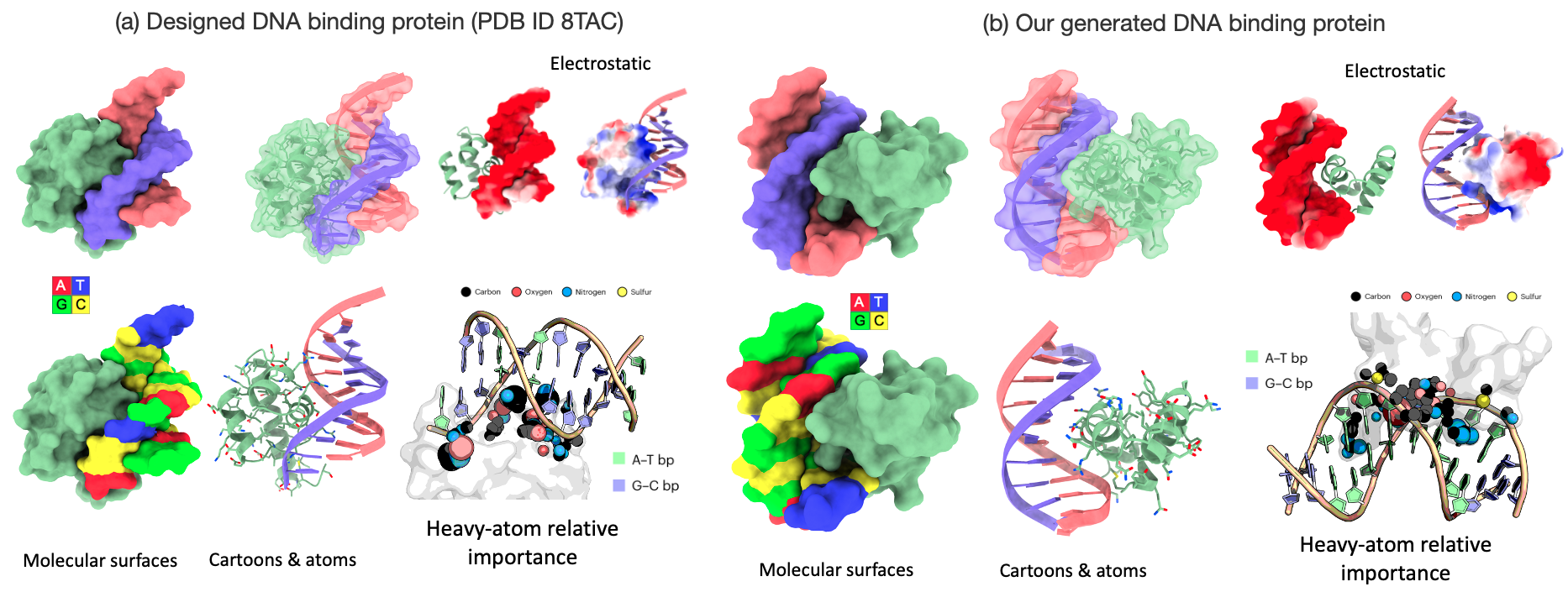}
    \caption{Generated binder example to specific DNA Chains C (ACCTGACGCGA) and D (TTCGCGTCAGG) in comparison with designed DNA binding protein (PDB ID 8TAC).}
    \label{fig:8TAC}
\end{figure}

During inference, DNA embeddings $\mathbf{H}_d^{(1)}$ and $\mathbf{H}_d^{(2)}$ are passed to $M_p$, where the generation of the protein sequence begins with a fixed token, ``M'' as explained previously. By fixing the ``M'' token, the model can focus on generating the rest of the protein sequence while keeping the DNA context in mind. Figure \ref{fig:8TAC} illustrates an example in which the designed protein interacts with DNA chains \(X_d^{(1)} = (\text{ACCTGACGCGA})\) and \(X_d^{(2)} = (\text{TTCGCGTCAGG})\). The model generates a protein sequence \(\hat{X}_p\) leveraging DNA-informed representations \(\mathbf{H}_d^{(1)'}\) and \(\mathbf{H}_d^{(2)'}\), enforcing sequence-specific interactions. This generated binder is compared against a reference DNA-binding protein (PDB ID: 8TAC).

% This token serves as a controlled starting point for generating the protein sequence. The ``M'' token represents a generic initialization that anchors the generation process, ensuring that the protein sequence starts from a valid state. 

%Figure.\ref{fig:8TAC} illustrates a protein binder generated using our approach compared to a bonder designed computationally and released in 2024-08-28 (PDb ID 8TAC\footnote{\url{www.rcsb.org/structure/8TAC}}). One can appreciate how the generated protein (right panel (b)) binds the DNA structure which corresponds to the pairs of sequences of 'ACCTGACGCGA' and 'TTCGCGTCAGG'. The figure provides protein-DNA binding specificity and Heavy-atom relative importance for both binders using DeepPBS server\footnote{\url{https://deeppbs.usc.edu}} as in \cite{Glasscock2024}.     

\section{Experimental Evaluation}
\label{sec:experiments}

In this section, we report experimental evaluation performances of our Prot42-L model and fine-tuned variants on \textbf{1)} Prediction of protein function, prediction of subcellular localization, prediction of structure, prediction of protein-protein interaction, and prediction of protein-ligand interaction \cite{xu2023peer}; \textbf{2)} Protein binder generation for different challenging targets taken form literature \cite{watson2023novo} and \cite{zambaldi2024novo}; and \textbf{3)} DNA Sequence-specific binders generation on targets taken from literature \cite{Glasscock2024}. While the last two experiments focus on generation to help computationally design binders, the first experiment show state-of-the-art results on downstream prediction tasks.   

\subsection{Evaluation on PEER Benchmarks}

In this section, we present the evaluation of our protein foundational model, Prot42, on the PEER benchmark. The PEER benchmark provides a comprehensive and multitask evaluation framework for \textbf{P}rotein s\textbf{E}quence und\textbf{ER}standing (PEER), covering a diverse set of downstream tasks. These tasks encompass prediction of protein function, prediction of subcellular localization, prediction of structure, prediction of protein-protein interaction, and prediction of protein-ligand interaction \cite{xu2023peer}.
To assess the performance of Prot42, we compare it with existing models on the PEER benchmark leaderboard. We report the performance of our model on all 14 benchmark tasks and provide a comparative analysis against the top five models of the PEER benchmark. Our results highlight the generalization and effectiveness of Prot42 in different protein modeling tasks, demonstrating its ability to learn meaningful representations for various biological contexts.

\begin{table}[ht!]
\centering
\scriptsize
\setlength{\tabcolsep}{2pt}
\resizebox{\textwidth}{!}{%
\begin{tabular}{l|c|c|c|c|c|c|c|c}
\textbf{PEER tasks} \cite{xu2023peer} & \textbf{Metric} & \textbf{CNN} & \textbf{CNN+Contact} & \textbf{CNN+SSP} & \textbf{ESM1b} & \textbf{ProtBert} & \textbf{Transformer} & \textbf{Prot42} \\
\shline
\multicolumn{9}{c}{\cellcolor{gray!25}\textbf{Protein Function Prediction}} \\
%\hline
Fluorescence Prediction & Accuracy & 0.682 & 0.680 & 0.683 & 0.679 & 0.679 & 0.643 & \textbf{0.685} \\
%\hline
Stability Prediction & Spearman's Rho & 0.637 & 0.661 & 0.695 & 0.694 & \textbf{0.771} & 0.649 & 0.756 \\
%\hline
Beta-lactamase Activity Prediction & Spearman's Rho & 0.781 & 0.835 & 0.811 & 0.839 & 0.731 & 0.261 & \textbf{0.876} \\
%\shline
Solubility Prediction & Accuracy & 0.644 & 0.706 & 0.699 & 0.702 & 0.682 & 0.701 & \textbf{0.752} \\
\shline
\multicolumn{9}{c}{\cellcolor{gray!25}\textbf{Protein Subcellular Localization Prediction}} \\
%\hline
Subcellular Localization & Accuracy & 0.587 & 0.591 & 0.566 & \textbf{0.781} & 0.765 & 0.560 & 0.780 \\
%\hline
Binary Localization Prediction & Accuracy & 0.827 & 0.827 & 0.818 & 0.924 & 0.913 & 0.757 & \textbf{0.936} \\
\shline
\multicolumn{9}{c}{\cellcolor{gray!25}\textbf{Protein Structure Prediction}} \\
\hline
Contact Prediction & L/5 precision & 0.100 & - & 0.057 & \textbf{0.458} & 0.397 & 0.175 & 0.279 \\
%\hline
Fold Classification & Accuracy & 0.109 & 0.111 & 0.117 & 0.282 & 0.169 & 0.085 & \textbf{0.344} \\
%\hline
Secondary Structure Prediction & Accuracy & 0.661 & 0.661 & - & \textbf{0.827} & 0.822 & 0.596 & 0.759 \\
\shline
\multicolumn{9}{c}{\cellcolor{gray!25}\textbf{Protein-Protein Interaction Prediction}} \\
%\hline
Yeast Protein Interaction & Accuracy & 0.551 & 0.545 & 0.541 & 0.570 & \textbf{0.637} & 0.541 & 0.604 \\
%\hline
Human Protein Interaction & Accuracy & 0.626 & 0.651 & 0.664 & \textbf{0.782} & 0.773 & 0.596 & 0.738 \\
%\hline
PPI Affinity Prediction & RMSE & 2.796 & \textbf{1.732} & 2.270 & 2.281 & 2.195 & 2.499 & 2.735 \\
\shline
\multicolumn{9}{c}{\cellcolor{gray!25}\textbf{Protein-Ligand Interaction Prediction}} \\
%\hline
Protein-Ligand Interaction (PLI) & RMSE & 1.376 & 1.328 & 1.295 & 1.559 & 1.562 & 1.455 & \textbf{1.250} \\
%\hline
PLI Affinity BindingDB & RMSE & 1.497 & 1.501 & 1.481 & 1.556 & 1.549 & 1.566 & \textbf{1.350} \\
\end{tabular}%
}
%\end{table}
\caption{Performance comparison on various protein foundational tasks across different models. \textbf{Bold} indicates the best performance per task (Results on existing techniques are reported as they are from the PEER Leaderboard\protect\footnote{\protect\url{https://torchprotein.ai/benchmark}}).}

\label{tab:Prot42_PEER_benchmark}
\end{table}

Our proposed model, Prot42, was evaluated on these tasks to benchmark its predictive capabilities comprehensively. Prot42 consistently demonstrated superior or highly competitive performance compared to baseline models, highlighting its robustness and generalizability. Specifically, \textbf{1)} Prot42 achieved superior performance in predicting stability, solubility, and beta-lactamase activity, highlighting its potential in high-resolution protein engineering tasks; \textbf{2)} In localization tasks (Binary and Subcellular), Prot42 performance rivaled established models, indicating its utility in functional annotation; \textbf{3)} Structural prediction tasks (Contact, Fold, Secondary Structure) showed strong results, reflecting Prot42's ability to capture structural nuances; \textbf{4)} For Protein-Protein and Protein-Ligand Interaction predictions, Prot42 demonstrated high precision and reliability, confirming its suitability for complex biological interaction modeling and pharmaceutical applications. In the protein-ligand interaction predictions, we utilized Chem42 for generating chemical embeddings. We also performed comparative analyses with ChemBert as an alternative chemical representation model, where we still outperformed existing methodologies with performance metrics approaching those achieved with Chem42 \cite{Chem42}. Overall, Prot42 demonstrates excellent potential across diverse biological prediction scenarios, highlighting its utility for advanced research and practical applications in bioengineering and pharmaceutical sciences.

\subsection{Protein Binders generation}
\label{sec:protein_binders_generation}
To rigorously assess the effectiveness of Prot42 for protein binder generation, we compared our model against AlphaProteo, a state-of-the-art model specifically designed for protein binder prediction. We selected AlphaProteo as our benchmark due to its established performance in generating high-affinity binders for clinically relevant targets. %\textcolor{red}{[This section need to mention more RFdiffusion \cite{watson2023novo} in addition to AlphaProteo \cite{zambaldi2024novo} -- Amaan]}.

\begin{itemize}
    \item 2wh6\footnote{\url{https://www.rcsb.org/structure/2WH6}} (anti-apoptotic BHRF1): The BHRF1 protein, encoded by the Epstein-Barr virus (EBV), is a viral homolog of the Bcl-2 family and functions as an anti-apoptotic factor. The crystal structure (PDB ID: 2WH6) reveals a characteristic Bcl-2-like fold composed of multiple $\alpha$-helices forming a hydrophobic groove, which facilitates the binding and sequestration of pro-apoptotic proteins such as Bim and Bak. By inhibiting mitochondrial outer membrane permeabilization (MOMP), BHRF1 enhances cell survival, contributing to viral persistence and immune evasion.

    \item 6m0j\footnote{\url{https://www.rcsb.org/structure/6m0j}} (SARS-CoV-2 spike receptor-binding domain bound): The receptor-binding domain (RBD) of the SARS-CoV-2 spike (\textit{S}) protein is a critical region that mediates viral entry into host cells by interacting with the human angiotensin-converting enzyme 2 (ACE2) receptor. The RBD, located within the S1 subunit, undergoes conformational changes between ``up'' and ``down'' states, regulating its accessibility for ACE2 binding. This domain contains key residues essential for receptor recognition and viral attachment, making it a primary target for neutralizing antibodies and vaccine development. Mutations in the RBD can influence viral transmissibility, immune escape, and therapeutic efficacy.

    \item 3di3\footnote{\url{https://www.rcsb.org/structure/3di3}} (glycosylated human interleukin-7 receptor alpha ectodomain): The glycosylated human interleukin-7 receptor alpha (\textit{IL-7R$\alpha$}) is a transmembrane glycoprotein that plays a crucial role in lymphocyte development and homeostasis. It is a component of the heterodimeric IL-7 and thymic stromal lymphopoietin (TSLP) receptor complexes. N-linked glycosylation of \textit{IL-7R$\alpha$} is essential for proper folding, stability, and receptor-ligand interactions, influencing signal transduction pathways that regulate T-cell survival and proliferation. Dysregulation of \textit{IL-7R$\alpha$} expression or glycosylation has been implicated in immune deficiencies and leukemogenesis.

    \item 5o45\footnote{\url{https://www.rcsb.org/structure/5O45}} (PD-L1): Programmed death-ligand 1 (\textit{PD-L1}), also known as B7-H1 or CD274, is a transmembrane protein expressed on antigen-presenting cells and various tumor cells. It plays a crucial role in immune regulation by binding to its receptor, programmed death-1 (\textit{PD-1}), on T cells, leading to immune suppression and tolerance. This interaction is a key mechanism in immune evasion by tumors, making \textit{PD-L1} a significant target for cancer immunotherapy. Inhibiting \textit{PD-L1} with immune checkpoint inhibitors enhances T cell activity, restoring anti-tumor immune responses.

    \item 1www\footnote{\url{https://www.rcsb.org/structure/1WWW}} (Tropomyosin receptor kinase A): The Tropomyosin receptor kinase A (\textit{TrkA}), also known as neurotrophic receptor tyrosine kinase 1 (\textit{NTRK1}), is a receptor tyrosine kinase that plays a crucial role in neuronal development, differentiation, and survival. It is activated by its ligand, nerve growth factor (NGF), leading to autophosphorylation and downstream signaling through the MAPK, PI3K-Akt, and PLC$\gamma$ pathways. \textit{TrkA} mutations and fusions are associated with oncogenic signaling in various cancers, making it a therapeutic target for selective tyrosine kinase inhibitors.

    \item 1bj1\footnote{\url{https://www.rcsb.org/structure/1BJ1}} (vascular endothelial growth factor A): Vascular endothelial growth factor A (\textit{VEGF-A}) is a potent angiogenic cytokine involved in the regulation of vascular growth and permeability. It exerts its biological functions by binding to VEGF receptors (primarily VEGFR-2) on endothelial cells, initiating downstream signaling cascades that promote endothelial cell proliferation, migration, and survival. The crystal structure (PDB ID: 1BJ1) reveals a homodimeric protein with a characteristic cystine-knot motif, essential for receptor binding. Dysregulated \textit{VEGF-A} expression contributes to pathological angiogenesis in cancer, diabetic retinopathy, and other diseases, making it a major target for anti-angiogenic therapies.

    \item 1tnf\footnote{\url{https://www.rcsb.org/structure/1TNF}} (TNF$\alpha$): Tumor Necrosis Factor alpha (TNF$\alpha$) is a pro-inflammatory cytokine primarily produced by macrophages and other immune cells. It plays a crucial role in immune responses, inflammation, and apoptosis. TNF$\alpha$ exerts its effects by binding to TNF receptors (TNFR1 and TNFR2), activating signaling pathways such as NF-$\kappa$B and MAPK, which regulate cell survival, differentiation, and immune modulation. Dysregulation of TNF$\alpha$ is implicated in various inflammatory diseases, including rheumatoid arthritis, Crohn's disease, and cancer.
\end{itemize}

\paragraph{Experimental Setting}
For fine-tuning Prot42 to generate protein binders, we used the STRING database~\cite{szklarczyk2021string}, a comprehensive resource for protein-protein interactions. STRING integrates experimental data, computational predictions, and text mining to provide confidence scores for protein interactions across multiple organisms. We applied stringent filtering criteria to extract high-quality binding pairs.
\begin{enumerate}
    \item Only interaction pairs with confidence scores $\geq 90\%$ were selected, ensuring a high reliability of the binding relationships.
    \item Sequences were limited to $\leq 250$ amino acids in length to focus on manageable, single-domain binding proteins, such that $n, m \leq 250$.
    \item We excluded redundant pairs to prevent overfitting and ensure diversity in the training dataset.
\end{enumerate}

After filtering, our final dataset consisted of 74,066 protein-protein interaction pairs from the STRING database. We divided this into a training set $\mathcal{D}_{pb}^{train}$ with 59,252 samples and a validation set $\mathcal{D}_{pb}^{val}$ with 14,814 samples (approximately 80\%/20\% split). The validation set $\mathcal{D}_{pb}^{val}$ was primarily used to track training loss and optimize the model's learning process. 
\begin{table}[ht!]
\centering
\footnotesize
\begin{tabular}{l|c}
\textbf{Hyperparameter} & \textbf{Value} \\
\hline
Optimizer & AdamW \\
Learning rate & $3 \times 10^{-5}$ \\
Learning rate schedule & Cosine decay with linear warm-up \\
Warm-up steps & 1000 \\
Minimum learning rate & $1 \times 10^{-6}$ \\
Momentum parameters ($\beta_1$, $\beta_2$) & 0.9, 0.999 \\
Weight decay & 0.01 \\
Batch size & 64 \\
Training epochs & 5 \\
Maximum sequence length & 512 tokens (including both target and binder) \\
Gradient clipping & 1.0 \\
Precision & Mixed precision (fp16) \\
\end{tabular}
\caption{Hyperparameters used for fine-tuning the protein binder generation models.}
\label{table:training_hyperparams}
\end{table}

We utilized the same target proteins and binding hotspots that were validated in AlphaProteo studies. These targets represent diverse therapeutic domains: BHRF1 (an Epstein-Barr virus protein that promotes cancer by inhibiting apoptosis), Sc2RBD (SARS-CoV-2 receptor binding domain, critical for viral entry into host cells), IL-7R$\alpha$ (Interleukin-7 receptor alpha, implicated in leukemia and HIV pathogenesis), PD-L1 (Programmed death-ligand 1, a key target in cancer immunotherapy), TrkA (Tropomyosin receptor kinase A, involved in chronic pain and autoimmune conditions), VEGF-A (Vascular endothelial growth factor, critical in cancer and eye disease progression), and TNF$\alpha$ (Tumor necrosis factor alpha, a central mediator in autoimmune disease).

Unlike AlphaProteo, which directly incorporates binding site information in its generation process, Prot42 operates solely on sequence inputs. To address this difference, we implemented a two-stage approach: (1) a generation phase, where for each target, we generated 500 candidate binder sequences using Prot42-L with the 8K context window; and (2) a filtering phase, where we modeled the 3D structure of each target-binder complex using Boltz-1 and retained only sequences positioned within 6-8 Å of all identified binding hotspots.
The diversity of our sampling approach yielded a rich exploration of the sequence space, producing binding candidates with varying properties, predicted affinities and ranked using computational validation metrics discussed in evaluation metrics paragraph (\ref{para:eval_metrics})

\paragraph{Evaluation Metrics}\label{para:eval_metrics}

We established a comprehensive evaluation framework to assess the quality of generated protein binders, focusing primarily on binding affinity by estimating the Dissociation Constant ($K_d$) and interface characteristics. Our analysis utilizes Prodigy (PROtein binDIng enerGY prediction) \cite{prodigy}, a well-established tool for quantifying protein-protein interactions based on structural features.

For each generated binder, we first predicted the 3D structure of the target-binder complex using Boltz-1, a state-of-the-art structure prediction model specifically designed for multimeric protein complexes. The resulting complexes were then analyzed using the following metrics:

\begin{enumerate}
    \item \textbf{Dissociation Constant ($K_d$):}  
    A concentration‐based measure of binding affinity in molar units, defined at equilibrium by  
    \[
      K_d \;=\; \frac{[P]\,[L]}{[PL]}
    \]
    where $[P]$, $[L]$, and $[PL]$ are the equilibrium concentrations of free protein, free ligand, and protein–ligand complex, respectively.  
    Lower $K_d$ values correspond to tighter binding (e.g.\ $K_d<10^{-9}\,\mathrm{M}$ indicates high affinity).

    \item \textbf{Binding Free Energy ($\Delta G$):}  
    The thermodynamic potential of binding, estimated in kcal/mol (e.g.\ via Prodigy’s empirical scoring function).  It is related to $K_d$ by
    \begin{equation}
      \Delta G \;=\; RT \,\ln K_d
    \end{equation}
    where $R$ is the gas constant and $T$ the absolute temperature (typically 310 K).  
    More negative $\Delta G$ denotes stronger binding, with values below about –9.0 kcal/mol generally indicating high‐affinity interactions.
    
    \item \textbf{Interface Characteristics:} 
    Several interface properties were quantified, including interface area (Å$^2$), the number of residue pairs at the binding interface (within 5.5Å), and the distribution of polar and non-polar contacts across the interface.
\end{enumerate}
% TODO : Update the plot to remove IL7A. 
\begin{figure}[ht!]
    \centering
    \includegraphics[width=\linewidth]{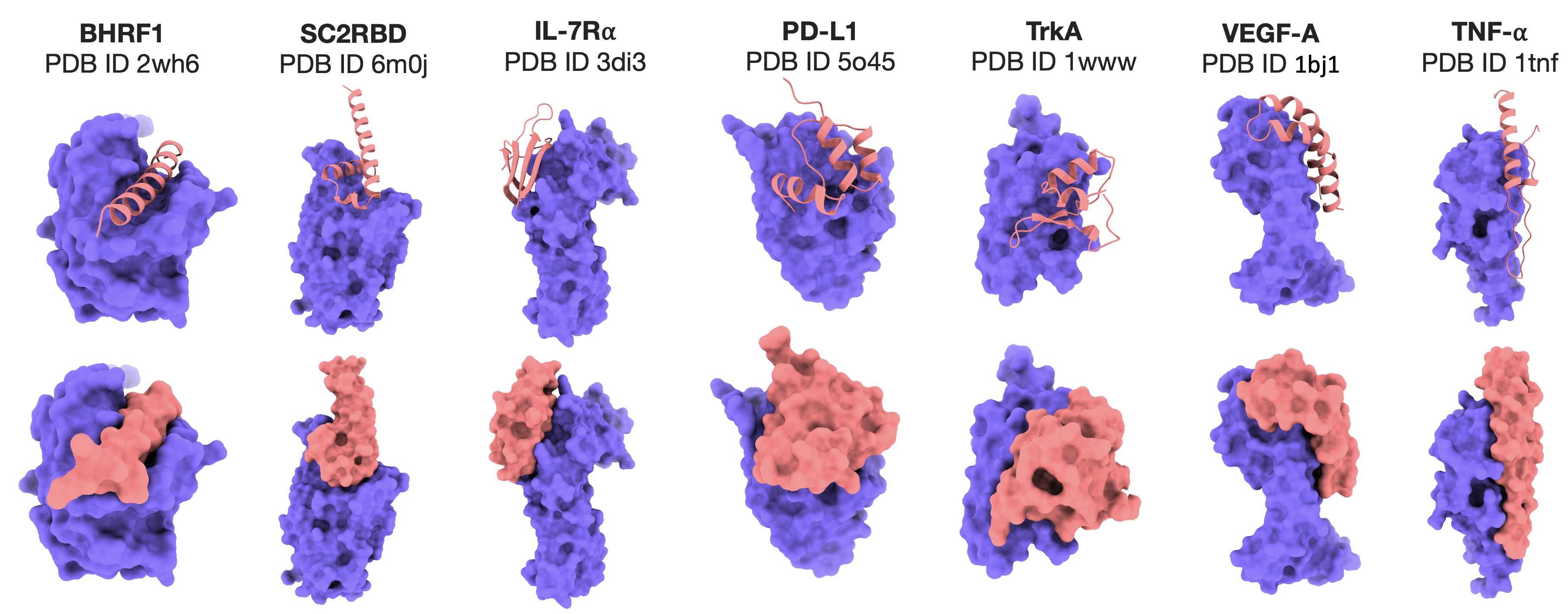}
    \caption{Examples of generated protein binders (green molecular surfaces - structures in different colors) to target protein sequences provided in the PDB IDs.}
    \label{fig:enter-label}
\end{figure}

\begin{figure}[ht!]
\centering
\includegraphics[width=1\textwidth]{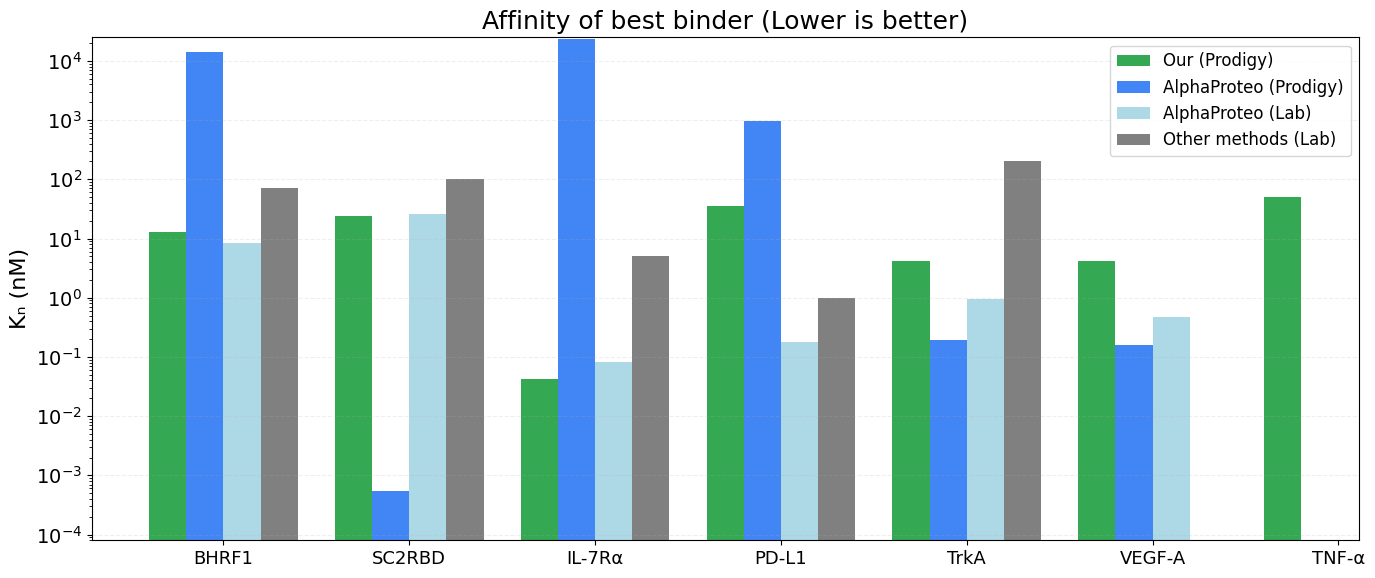}
\caption{\textbf{Comparison of dissociation constants ($K_d$, nM) for top binders against seven therapeutic targets.} Green bars represent Prot42 Prodigy predictions; dark‐blue bars, AlphaProteo in silico Prodigy predictions; light‐blue bars, AlphaProteo laboratory measurements; and gray bars, other design methods. Lower $K_d$ denotes tighter binding. All values are plotted on a base-10 logarithmic axis.}

\label{fig:binding_affinity}
\end{figure}

\paragraph{Binding Affinity Comparison}
Figure \ref{fig:binding_affinity} compares the dissociation constants ($K_d$, in nM) for the strongest binders against BHRF1, SC2RBD, IL-7R$\alpha$, PD-L1, TrkA, VEGF-A and TNF-α.  The bars show the Prodigy predicted values of our model $K_d$, AlphaProteo Prodigy predictions in silico, AlphaProteo laboratory measurements, and benchmark values from other design methods.  Lower $K_d$ values indicate tighter binding.  All $K_d$ values were obtained by converting Prodigy ($\Delta G$) estimates to dissociation constants and are plotted on a base-10 logarithmic axis.

Our results demonstrate that Prot42 consistently generated binders with strong predicted affinities across all tested targets. For IL-7R$\alpha$, Prot42 achieved a dissociation constant of 0.043 nM, markedly improving on AlphaProteo’s in silico prediction of 23,000 nM. Against PD-L1, Prot42 produced a $K_d$ of 35 nM compared to AlphaProteo’s 980 nM. Prot42 also delivered low-nanomolar binding for TrkA (4.1 nM vs. 0.19 nM) and VEGF-A (4.2 nM vs. 0.16 nM), underscoring its robust performance across both immune and growth-factor targets.

\begin{figure}[ht!]
\centering
\includegraphics[width=1\textwidth]{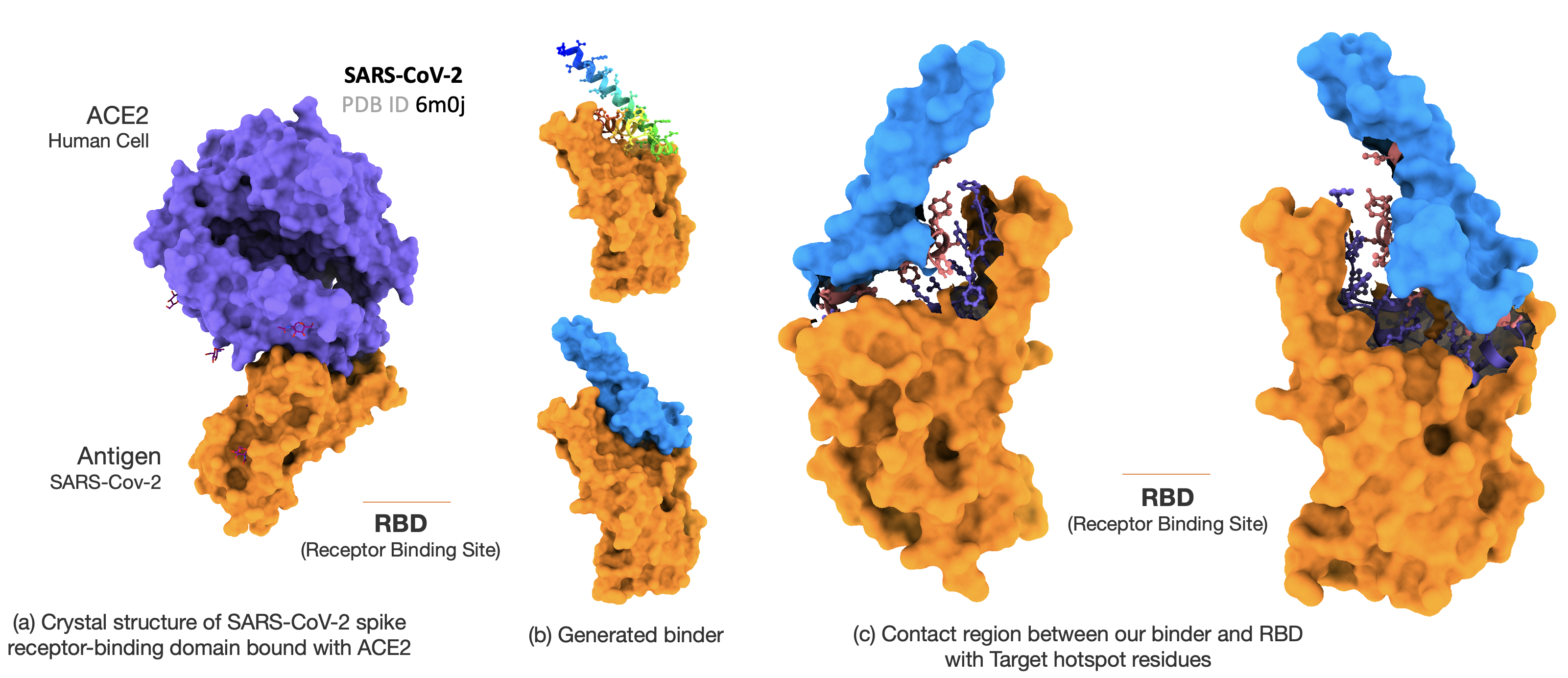}
\caption{Zooming on the generated protein binder for the RBD of SARS-CoV-2 (PDB ID 6m0j) with the contact region between both proteins.}
\label{fig:6m0j}
\end{figure}

Figure \ref{fig:6m0j} illustrates our generated top-ranked binder for the receptor-binding domain (RBD) of the SARS-CoV-2 spike glycoprotein (PDB ID 6m0j). For this target, we used residues 333-526 of chain E, which encompass the entire RBD domain responsible for host cell recognition. We conditioned our Prot42 model to generate a binding protein with a minimum length of 50 amino acids to ensure sufficient interface area for stable interaction. The generated binder was specifically filtered from a candidate pool based on proximity (6-8 Å) to the critical hotspot residues E485, E489, E494, E500 and E505 which constitute the receptor-binding motif (RBM). These residues are known to form critical interactions with the peptidase domain of human angiotensin-converting enzyme 2 (ACE2). By targeting this specific epitope, our designed binder is predicted to function as a competitive inhibitor, effectively blocking the interaction between the SARS-CoV-2 spike protein and ACE2, thus potentially neutralizing viral entry into host cells.

\subsection{DNA Sequence-specific Binders Generation}

This task focuses on designing proteins that are capable of binding to target DNA sequences. To achieve this, we used the 2010 Protein-DNA Interface Database (PDIdb) 2010 dataset, as outlined in \cite{norambuena2010protein}. The dataset contains 922 unique DNA-protein pairs, which were used as a basis for our model. For the evaluation of our DNA-Protein model, we extracted DNA segments from various PDB structures, including 1TUP, 1BC8(56), 1YO5(57), 1L3L(44), 2O4A(58), 1OCT(59), 1A1F(60), and 1JJ6(61). 

\begin{figure}[ht!]
    \centering
    \includegraphics[width=\linewidth]{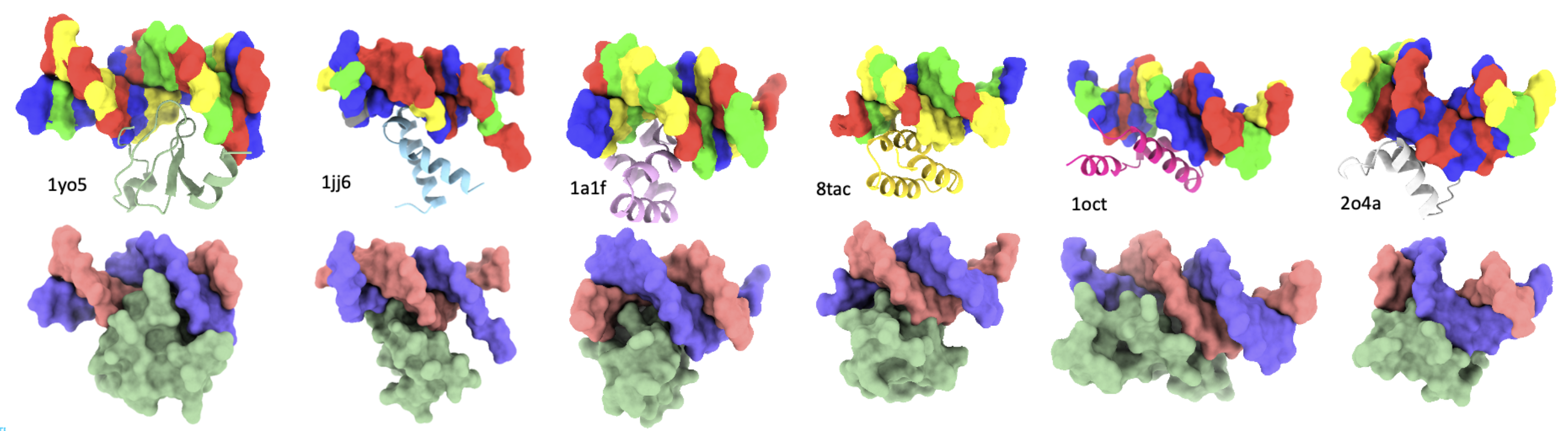}
    \caption{Examples of generated protein binders (green molecular surfaces - structures in different colors) to DNA sequences provided in the PDB IDs: 1yo5, 1jj6, 1a1f, 8tac, 1oct and 2o4a.}
    \label{fig:DNA-binding}
\end{figure}

In addition to this, we used DeepPBS \cite{mitra2023deeppbs} to assess the binding specificity of the generated proteins. DeepPBS is a model designed to capture the physicochemical and geometric contexts of protein-DNA interactions. It predicts binding specificity, represented as a position weight matrix (PWM), based on a given protein-DNA structure. High specificity interactions are expected to maximize affinity across various DNA base possibilities, and the importance scores generated by DeepPBS for these interactions correlate with the corresponding binding affinities. A PWM is defined as an \( N \times 4 \) matrix, where \( N \) is the length of the DNA sequence of interest, and the four positions correspond to the four DNA bases: adenine (A), cytosine (C), guanine (G), and thymine (T). Each column of the PWM represents the probabilities of the four bases occurring at that specific position within the sequence. Figure \ref{fig:DNA-binding} illustrates several examples of protein binders to different target DNA sequences as they are given in PDB ID 1yo5, 1jj6, 1a1f, 8tac, 1oct and 2o4a. The figure shows both structures in cartoon representation and molecular surfaces to highlight the ability of our approach to account for the structures when generating the protein binders.           

\section{Conclusion}
\label{sec:Conclusion}

In this work, we introduced Prot42, a pioneering family of Protein Language Models (pLMs) designed to generate high-affinity protein binders solely from sequence information. Prot42 demonstrates that the latent evolutionary and functional information embedded within raw protein sequences is sufficient to drive highly accurate, large-scale protein generation, without requiring structural input. By leveraging an auto-regressive, decoder-only architecture, our models overcome traditional constraints on sequence length and generative fidelity, enabling the capture of complex long-range dependencies and intricate multi-domain relationships—key for designing functional protein binders with unprecedented precision.
Through extensive evaluations, Prot42 has proven its ability to synthesize sequence-specific Protein and DNA binding proteins, significantly accelerating the binder design process while reducing dependence on labor-intensive experimental discovery. Unlike structure-dependent approaches, our method highlights the untapped generative potential of sequence-based modeling, demonstrating that protein functionality can be accurately inferred, optimized, and expanded through language modeling alone. Looking ahead, we plan to validate Prot42-generated binders experimentally, complementing computational assessments with real-world functional tests. This step will solidify the model’s utility in practical applications and refine its predictive accuracy, bridging the gap between AI-driven sequence generation and experimental biotechnology. By continuing to enhance sequence-based generative modeling, Prot42 paves the way for scalable, data-driven protein engineering, unlocking new frontiers in synthetic biology and therapeutic development.

\section*{Acknowledgment}
The authors are thankful to MIT Jameel Clinic for releasing Boltz-1, an open-source model designed to accurately model complex biomolecular interactions. All structures of our protein binders in complex with targets are produced using Boltz-1. All protein visualizations reported in this paper are obtained using the UCSF ChimeraX software.   

\bibliographystyle{plainnat}
\bibliography{main}

\end{document}